\documentclass{ws-procs9x6}

\def\theequation{\thesection.\arabic{equation}}
\newcommand{\cO}{\mathcal{O}}
\newcommand{\cN}{\mathcal{N}}
\newcommand{\cF}{\mathcal{F}}
\newcommand{\sPP}{\mathrm{I}\kern -1.6pt \mathrm{P}}
\newcommand{\sR}{\mathrm{I}\kern -1.6pt \mathrm{R}}
\newcommand{\tQ}{\tilde{Q}}
\newcommand{\tq}{\tilde{q}}
\newcommand{\La}{\Lambda}
\newcommand{\U}{\mathrm{U}}
\newcommand{\1}{1\kern -3pt \mathrm{l}}
\newcommand{\tr}{{\rm tr}}
\newcommand{\D}{{\rm d}}

\def\vev#1{ \langle {#1} \rangle }
\def\bvev#1{ \bigg\langle {#1} \bigg\rangle }
\def\vevS{ \bigg|_{\vev{S}}  }
\def\sumN{ \sum_{i=1}^{N} }
\def\sumjN{ \sum_{j=1}^{N} }
\def\sumI{ \sum_{I=1}^{N_f} }
\def\prodN{ \prod_{i=1}^{N} }
\def\gs{ g_{s} }
\def\free{ F_{\rm s}(S) }
\def\fdisk{ F_{\rm d} (S) }
\def\frp{ F_{\rm rp} (S) }
\def\oms{ \omega_{\rm s} }

\newcommand{\beq}{\begin{equation}}
\newcommand{\eeq}{\end{equation}}
\newcommand{\bea}{\begin{eqnarray}}
\newcommand{\eea}{\end{eqnarray}}

\newcommand{\ts}{\textstyle}
\newcommand{\non}{\nonumber}

\def\draftnote{}

\begin{document}

\title{
\vspace{-5\baselineskip}
\begingroup
\footnotesize\normalfont\raggedleft
\lowercase{\sf hep-th/0401108} \\ 
BRX-TH-530\\
BOW-PH-130\\
CERN-PH-TH/2004-005\\
\vspace{\baselineskip}
\endgroup
Matrix models and $\cN=2$ gauge theory\footnote{
\uppercase{T}o appear in the \uppercase{P}roceedings of the 3rd 
\uppercase{S}ymposium on \uppercase{Q}uantum
\uppercase{T}heory and \uppercase{S}ymmetries (\uppercase{QTS}3),
\uppercase{C}incinnati, \uppercase{O}hio,
10--14 \uppercase{S}ept 2003 ---
\copyright\
\uppercase{W}orld \uppercase{S}cientific.}
}

\author{S.~G.~NACULICH\footnote{\uppercase{R}esearch partially
supported by the \uppercase{NSF} under grant \uppercase{PHY}-0140281.}}

\address{Department of Physics\\
Bowdoin College\\ 
Brunswick, ME 04011 USA\\ 
E-mail: naculich@bowdoin.edu}

\author{H.~J.~SCHNITZER\footnote{\uppercase{R}esearch partially
supported by the \uppercase{DOE} under 
grant \uppercase{DE--FG02--92ER40706}.} ~and
N.~WYLLARD\footnote{\uppercase{R}esearch 
supported by the \uppercase{DOE} under 
grant \uppercase{DE--FG02--92ER40706}.
\uppercase{C}urrent address:  
\uppercase{D}epartment of \uppercase{P}hysics, 
\uppercase{CERN}, 
\uppercase{T}heory \uppercase{D}ivision,
\uppercase{CH}-1211, 
\uppercase{G}eneva 23, 
\uppercase{S}witzerland.}}

\address{Martin Fisher School of Physics\\ 
Brandeis University\\
Waltham, MA 02454 USA\\
E-mail: schnitzer@brandeis.edu, wyllard@cern.ch} 

\maketitle

\abstracts{
We describe how the ingredients and results 
of the Seiberg-Witten solution to $\cN=2$ 
supersymmetric $\U(N)$ gauge theory
may be obtained from a matrix model.
}

Dijkgraaf and Vafa discovered that the 
non-perturbative effective superpotential 
for certain $d=4$ $\cN=1$ supersymmetric gauge theories 
can be obtained by calculating planar diagrams 
in a related gauged matrix 
model\cite{Dijkgraaf:2002a,Dijkgraaf:2002b,Cachazo:2002}
(for a more complete list of references, see Ref.~\refcite{Argurio:2003}).
In this talk, we will show that matrix models 
can also be used to obtain all the
ingredients and results of the 
Seiberg-Witten solution\cite{Seiberg:1994}
of certain  $\cN=2$ supersymmetric gauge theories,
specifically $\U(N)$ theories without matter,
or with matter in fundamental, symmetric, or antisymmetric
representations\cite{Dijkgraaf:2002a,Dijkgraaf:2002b,Naculich:2002,Klemm:2003,Naculich:2003a,Naculich:2003b}.

\section{$\cN=2$ supersymmetric gauge theory}

We will focus on the $\cN=2$ $\U(N)$ gauge theory
with $N_f$ hypermultiplets in the fundamental representation
to illustrate the matrix model approach,
indicating where differences occur in 
theories with symmetric or antisymmetric  hypermultiplets.
To apply the insights of Dijkgraaf-Vafa to an $\cN=2$ gauge theory,
one begins by expressing its field content 
in terms of $\cN=1$ superfields.
Let $\phi$ denote the adjoint $\cN=1$ chiral superfield
belonging to the $\cN=2$ vector multiplet,
and $q^I$ and $\tilde{q}_I$ the $\cN=1$ chiral superfields 
that comprise the $\cN=2$ hypermultiplets
transforming in the fundamental representation.
The $\cN=2$ theory has the superpotential 
\beq
\label{eq:super}
W_{\cN=2}  (\phi,q,\tq) =
\sumI \left[ \tq_I \phi\, q^I + m_I \tilde{q}_I q^I  \right] 
\eeq
where $m_I$ are the masses of the fundamentals.
The Coulomb branch of the moduli space of vacua
is characterized by an arbitrary diagonal vev for the scalar field in $\phi$, 
but we may select a specific (but generic) point 
$\phi = \mathrm{diag} ( e_1, \cdots, e_N )$,
at which the $\U(N)$  gauge group is broken to $\U(1)^N$,
by adding a perturbation to the superpotential
\beq
\label{eq:fullsuper}
W (\phi,q,\tq) = W_{\cN=2}  (\phi,q,\tq) + W_0 (\phi)
\eeq
where
$ W_0' (x)  = \alpha \prodN (x-e_i)$.
The perturbation breaks the supersymmetry to $\cN=1$,
but the full $\cN=2$ supersymmetry will be restored 
by sending $\alpha \to 0$ at the end of the matrix model calculation.

\setcounter{equation}{0}
\section{The perturbative matrix model}

Each $\cN=1$ chiral superfield described in the
previous section has an analog (denoted by a capital letter)
in the corresponding matrix model;
specifically, an $M {\times} M$ hermitian matrix $\Phi$,
and $M$-dimensional vectors $Q^I$ and $\tQ_I$.
(The analog of a symmetric or antisymmetric matter hypermultiplet
would be an $M \times M$ symmetric or antisymmetric matrix.)
The superpotential (\ref{eq:fullsuper}) is reinterpreted\cite{Dijkgraaf:2002a}
as the potential of the matrix model,
whose partition function is thus
\beq
\label{eq:partition}
Z = \frac{1}{\mathrm{vol}(G)}
\int \D\Phi \, \D Q^I \D \tQ_I
\exp \left( - \frac{W(\Phi,Q,\tQ)}{\gs} \right) 
\eeq
where $G$ is the unbroken matrix model gauge group,
and $\gs$ is a parameter that 
will be taken to zero in the planar limit $M \to \infty$.
The matrix integral (\ref{eq:partition}) 
can be evaluated perturbatively about an extremum 
\beq
\label{Phinought}
\Phi_0 = 
\begin{pmatrix} 
e_1 \1_{M_1}& 0& \cdots& 0 \cr
0& e_2 \1_{M_2}& \cdots& 0 \cr
\vdots& \vdots& \ddots& \vdots \cr
0& 0& \cdots&  e_N \1_{M_N}  
\end{pmatrix},
\qquad
(Q^I)_0 = 0,
\qquad
(\tQ_I)_0 = 0
\eeq
where the $M_i$ are arbitrary, subject to $\sumN M_i = M$.
The $\U(M)$ symmetry of the matrix model is broken 
by $\Phi_0$ to $G = \prod_{i=1}^N  \U(M_i)$. 
The residual gauge symmetry must be gauge-fixed, 
and ghosts introduced; for details, 
see Refs.~\refcite{Dijkgraaf:2002b}, \refcite{Naculich:2002}.
For the $\U(N)$ gauge theory with an antisymmetric 
representation\cite{Klemm:2003,Naculich:2003a,Naculich:2003b}
of mass $m$,
$\Phi_0$ must include an additional diagonal block $m \1_{M_0}$
and the antisymmetric matrix a corresponding block
$J$ (the symplectic unit), breaking the symmetry
of the matrix model to $\mathrm{Sp}(M_0) \times \prod_{i=1}^N  \U(M_i)$. 
The inclusion of the extra block for the antisymmetric case 
has been put in a broader context in Ref.~\refcite{Intriligator:2003}.

\subsection{Topological expansion} 

The connected diagrams of the perturbative expansion of (\ref{eq:partition})
may be organized, using standard 't Hooft double-line notation,
in a topological expansion characterized by the 
Euler characteristic $\chi$ of the surface 
in which the diagram is embedded
\beq
\label{eq:topol}
\log Z =  \sum_{\chi \le 2} \gs^{-\chi} F_{\chi} (S) ,
\qquad S_i = \gs M_i,
\qquad
\chi = 2 {-} 2g {-} h {-} q
\eeq
with $g$ the number of handles, 
$h$ the number of boundary components, 
and $q$ the number of crosscaps. 
We now take the large $M$ limit,
letting $M_i \to \infty$, $\gs \to 0$  with $S_i$  held fixed.
In this limit, the dominant contribution
$\free \equiv \gs^2 \log  Z \big|_{\rm sphere} $
arises from planar diagrams that can be drawn on a sphere.
Theories with fundamental representations
contain surfaces with boundaries;
the dominant such contribution 
$\fdisk \equiv \gs \log  Z \big|_{\rm disk} $
comes from planar diagrams on a disk.
Theories with symmetric or antisymmetric representations
contain nonorientable surfaces; 
the dominant nonorientable contribution
$\frp \equiv \gs \log  Z \big|_{\sR \sPP^2} $ 
comes from planar diagrams on 
$\sR \sPP^2$, a sphere with one crosscap.

\subsection{The effective superpotential} 

The values of $S_i$ in the matrix model, hitherto 
arbitrary, are determined by the 
extremization of the effective 
superpotential, given by\cite{Dijkgraaf:2002a,Argurio:2002,Ita:2002}
\beq
\label{eq:Weffdef}
W_{\rm eff} (S)
=-\bigg[\sumN \frac{\partial}{\partial S_i} \free + \fdisk + 4 \frp \bigg]
\eeq
in the case where the gauge group $\U(N)$
of the gauge theory is broken to $\U(1)^N$.
The resulting vevs $\vev{S_i}$ may be computed in an expansion in $\La$,
the scale in the matrix model.
For the $\U(N)$ theory with $N_f$ fundamentals,
the leading term is\cite{Naculich:2002}
\beq
\label{eq:Svev}
\vev{S_i} =
\alpha 
\frac{\prod_{I=1}^{N_f}(e_i+m_I)}{\prod_{j\neq i} (e_i-e_j)}
\La^{2N-N_f} + \cO(\La^{4N-2N_f})
\eeq
and the $\La^{4N-2N_f}$ term is
also computed in Ref.~\refcite{Naculich:2002}.

\subsection{Tadpole diagrams} 

The Seiberg-Witten solution 
of the $\cN=2$ gauge theory is expressed,
not in terms of the parameters $e_i$, 
but in terms of the renormalized order parameters $a_i$,
defined as the periods of the Seiberg-Witten differential\cite{Seiberg:1994}.
The matrix model prescription for computing $a_i$ was
presented and motivated 
in Ref.~\refcite{Naculich:2002}:
\beq
\label{eq:adef}
a_i  =  e_i + \bigg[ \sumjN \frac{\partial}{\partial S_j} 
\gs \vev{\tr\, \Psi_{ii} }_{\rm sphere} 
+   \vev{\tr\, \Psi_{ii} }_{\rm disk} 
+ 4 \vev{\tr\, \Psi_{ii} }_{\rm rp} 
 \bigg] \vevS
\eeq
where $\Psi_{ii}$ is the $i$th diagonal block of $\Phi-\Phi_0$,
and the vevs in Eq.~(\ref{eq:adef}) represent tadpole diagrams with
the specified topology.
Equation (\ref{eq:adef}) may be computed in 
an expansion in $\La$;
the $\La^{2N-N_f}$ contribution agrees\cite{Naculich:2002}
with the one-instanton relation 
between $a_i$ and $e_i$ computed in SW theory\cite{DHoker:1996}.

\subsection{Period matrix and prepotential} 

In Seiberg-Witten theory, 
the matrix $\tau_{ij}$ of gauge couplings of the unbroken 
$\U(1)^N$ gauge theory is given by\cite{Seiberg:1994,DHoker:1996}
\bea
\label{eq:SWtau}
\tau_{ij}(a) &=&  \frac{\partial^2 \cF(a)}{\partial a_i \partial a_j} \\
\cF(a) 
&=&  
\cF_{\rm pert} (a) 
 + \frac{\La^{2N-N_f}}{2 \pi i}  
\sum_i 
\frac{ \prod_{I=1}^{N_f} (a_i + m_I) } {\prod_{j \neq i} (a_i-a_j)^2}
+ \cO(\La^{4N-2N_f}) \non
\eea
where $\cF_{\rm pert}(a)$ is the perturbative prepotential
and the one-instanton prepotential is shown explicitly.

The matrix model prescription for the gauge coupling matrix
is\cite{Dijkgraaf:2002a}
\beq
\label{eq:matrixtau}
\tau_{ij} = \frac{1}{2\pi i}
\frac{\partial^2 \free}{\partial S_i \partial S_j}
\bigg|_{S = \vev{S} }.
\eeq
In Ref.~\refcite{Naculich:2002},
this quantity was computed 
and re-expressed in terms of $a_i$ using Eq.~(\ref{eq:adef}),
and was shown to agree with Eq.~(\ref{eq:SWtau})
to one-instanton accuracy.

For theories with symmetric or antisymmetric hypermultiplets,
the prescription (\ref{eq:matrixtau}) must be modified by
including relative signs among the various contributions to $\free$.
The justification for these signs, 
together with a prescription for computing $\tau_{ij}$,
was given in Refs.~\refcite{Naculich:2003a,Naculich:2003b}.
With this modification, 
and also the inclusion of the extra block for the case with
antisymmetric matter discussed above,
the matrix model calculation of $\tau_{ij}$ agrees with the 
SW calculation to one-instanton accuracy\cite{Naculich:2003b}.

\setcounter{equation}{0}
\section{SW curve and differential from the matrix model} 

In this section, we will indicate 
how the usual ingredients of the Seiberg-Witten approach,
the SW curve and SW differential, 
may be obtained from the matrix model
using saddle point methods.
In this approach, one introduces the trace of the resolvent 
\beq
\label{eq:resolvedef}
\omega(z) = \gs \bvev{\tr \left(\frac{1} {z - \Phi}\right) }
\eeq
which, like the free energy (\ref{eq:topol}),
may be expressed in terms of a topological expansion, 
with $\oms(z)$ the leading term in the large $M$ limit.
The saddle-point approximation to (\ref{eq:partition}) 
implies\cite{Dijkgraaf:2002a}
\beq
\oms^2(z) - W_0' (z) \, \oms(z) + {\ts \frac{1}{4}} f(z) = 0
\eeq
where $f(z)$ is a polynomial, given by 
\beq
\label{eq:fdef}
f(z) = 4 \gs \bvev{\tr\!\left(\frac{W_0'(z)-W_0'(\Phi)}{z-\Phi}\right) } .
\eeq
The polynomial $f(z)$ is determined by extremizing the effective 
superpotential (\ref{eq:Weffdef}); 
this may be done exactly\cite{Naculich:2002}
using Abel's theorem,
or perturbatively\cite{Naculich:2003b} using (\ref{eq:Svev}), as follows:
\bea
f(z)
&=& 4 \sum_i \frac{W_0'(z)}{z-e_i} \vev{S_i} + \cO(S^2)  \\
&= & 4 \, \alpha^2 \La^{2N-N_f} 
\bigg[ \prod_{I=1}^{N_f} (z+m_I) - \tilde{T}(z)  \prodN (z-e_i)   \bigg]
+ \cO(\La^{4N-2N_f}) \non
\eea
where $\tilde{T}(z)$ is the polynomial part of 
$\prod_{I=1}^{N_f} (z+m_I)/ \prodN (z-e_i) $.
Defining $y(z)  = - 2 \oms(z) + W_0' (z)$,  one obtains
\beq
\label{eq:hyper}
y^2 =  W_0' (z)^2 - f(z)
\eeq 
precisely the Seiberg-Witten curve for this theory\cite{DHoker:1996}
for the choice of moduli $e_i$ consistent with Eq.~(\ref{eq:adef}).
See Refs.~\refcite{Naculich:2002}, \refcite{Naculich:2003b} for details.

The Seiberg-Witten differential may 
also be obtained in the matrix model 
approach as\cite{Gopakumar:2002,Naculich:2002,Cachazo:2002,Naculich:2003a}
\beq
\lambda_{SW} 
= z \bigg[\sumN \frac{\partial}{\partial S_i} \, \oms 
+ \omega_{\rm d} + 4 \,\omega_{\rm rp} \bigg]
\, \D z. 
\eeq

The cubic Seiberg-Witten 
curve (and associated SW differential)
for the gauge theory with symmetric or antisymmetric matter hypermultiplets
may also be obtained\cite{Klemm:2003,Naculich:2003a,Naculich:2003b}
from the matrix model approach, 
using saddle-point methods, 
together with extremization of the effective superpotential.

\section*{Acknowledgments}
SGN would like to thank the organizers of the QTS3 conference
for the opportunity to present this work.

\end{document}